\documentclass[aps,prd,showpacs,nofootinbib,floats,floatfix,preprintnumbers
,groupedaddress,11pt]{revtex4}
\usepackage{comment}
\usepackage[titletoc,title]{appendix}
\usepackage{graphicx}
\usepackage{amsmath}
\usepackage{float}
\usepackage{subfigure}
\usepackage[usenames]{color}
\usepackage{amssymb}
\usepackage{epstopdf}
\newcommand{\bea}{\begin{aligned}}
\newcommand{\eea}{\end{aligned}}
\newcommand{\be}{\begin{equation}}
\newcommand{\ee}{\end{equation}}
\newcommand{\pr}{\partial}

\newcommand{\bse}{\begin{subequations}}
\newcommand{\ese}{\end{subequations}}

\newcommand{\km}{\mathrm{k}}
\newcommand{\mcM}{\mathcal{M}}
\usepackage{hyperref}
\hypersetup{
     colorlinks   = true,
     citecolor    = red,
     linkcolor    = blue,
     urlcolor     = blue,
}

\renewcommand{\v}[1]{\ensuremath{\mathbf{#1}}} 
\newcommand{\bmm}{\begin{multline}}
\newcommand{\emm}{\end{multline}}

\newcommand{\mi}{\mathrm{i}}
\numberwithin{equation}{section}
\begin{document}
\title{Ringing black holes are superradiant: The case of ultralight scalar fields}
\author{Rajesh Karmakar\footnote{rajesh018@iitg.ac.in}}
\author{Debaprasad Maity, \footnote{debu@iitg.ac.in}}
\affiliation{Department of Physics, 
Indian Institute of Technology, Guwahati, Assam 781039, India}
\pagenumbering{arabic}
\renewcommand{\thesection}{\arabic{section}}

\begin{abstract}
 
 Superradiance has been studied quite extensively in the context of static (charged) and rotating black hole spacetime. In this paper, we report for the first time that for a minimally coupled scalar field, the absorption cross-section of a Schwarzschild black hole in its ring down phase can be superradiant. Treating the scattered scalar field as axion, we further computed its observable effects on the rotation of the plane of polarization of photon, neutron dipole moment, and fermionic spin precession. All those observables turned out to have interesting time dependence induced by the ringing black hole background. Our present result opens up an intriguing possibility of observing the black hole merging phenomena through other fundamental fields.

\end{abstract}

\maketitle

\newpage
\section{Introduction}
Classical and quantum aspects of superradiance have been the subject of intense research in the context of black hole (BH) physics \cite{Brito:2015oca,Leite:2018mon,Dolan:2011ti,Cardoso:2019dte,bhbomb,Benone:2019all,Balakumar:2020gli}.
Though this fascinating phenomenon has not been observed yet, recent observation of gravitational waves by LIGO  and Virgo \cite{LIGOScientific:2016aoc, LIGOScientific:2016emj, LIGOScientific:2016vbw, LIGOScientific:2016vlm, LIGOScientific:2016sjg, LIGOScientific:2017ycc, LIGOScientific:2017vwq, LIGOScientific:2020stg, LIGOScientific:2016jlg,LIGOScientific:2020aai, LIGOScientific:2020zkf,LIGOScientific:2017vox} has paved an interesting possibility of observing such phenomena. LIGO has observed a large number of BH-BH or BH-neutron star merging phenomena which sets the beginning of a new era of gravitational-wave astronomy. After such observation, a surge of theoretical works has been devoted towards understanding the nature of spacetime in the strong gravity regime. In this parlance we ask the following question:{\em Along with the gravitational waves, are there any complementary observables that can shed further light on the nature of black holes?} Motivated by this question in this paper we shall study the scattering of a scalar field minimally coupled with the gravitational wave(ringing) black hole background.
 
 After two black holes merge, it undergoes different phases of evolution. The most significant one is the ring down phase through which the merged black hole settles down to its equilibrium states. During this phase, the black hole undergoes damped oscillation which has been extensively studied using the method of BH perturbation theory. Because of inherent dissipation, the oscillation frequency is quasi-normal. 
Quasi-normal mode (QNM) analysis of BHs has been studied extensively in the literature\cite{Cardoso:2003vt,Berti:2009kk,Flachi:2012nv,Nollert:1999ji,Frolov:2018ezx,Kokkotas:1999bd,Konoplya:2011qq}. In this paper, we study the wave scattering phenomena to see the response of such ringing black holes when interacting with the incoming scalar waves. Scattering of various fundamental fields has been widely studied in the static BH background \cite{vishvesh1970,Das:1996we,Unruh:1976fm,Crispino:2007qw,Benone:2014qaa}. However, understanding the ringing BHs interacting with the external waves not only sheds light on how black holes react with their environment, further, it could also lead to observable signatures which are complementary to the gravitational wave. Putting it in a different perspective, the present study has got interesting resemblance with the well-known "preheating" mechanism in inflationary cosmology where oscillating inflaton field in the vacuum imparts its energy to daughter fields through parametric resonance \cite{Kofman:1997yn}. This leads to the growth of outgoing flux of the daughter fields. The present study shows similar phenomena where the out going flux is enhanced compared to the incoming one after the interaction with the ringing black holes. In the present context, we call it superradiance which is quantified by the momentum-dependent negative absorption cross-section. 
As pointed out before, the flux of those resonant modes can act as complementary observables along with the gravitational waves during the ringing phase of the black holes.   
Hence, our present study opens up possibilities of understanding BH merging phenomena with other fundamental fields.

The present study will be confined to the ultralight scalar field, which can be identified as an axion or axion-like particle, which is of particular interest as a possible candidate for dark matter. The detectability of dark matter has been a pertinent issue and there have been several proposals and experiments \cite{MarrodanUndagoitia:2015veg, Leane:2020liq,Graham:2013gfa} dealing with the possible signature of the presence of dark matter. We have considered in this paper various observables which contain the signature of the oscillating scalar field arising due to the ringing background. Such a time-varying effect would be easier to detect.
Throughout our paper, we use GW background and ringing background interchangeably. 

We have organized the paper as follows. In Sec.\ref{metric}, we have described the background space-time in the ringing phase of the Schwarzschild black hole. For more quantitative details of the ringing background metric, one may refer to  Appendix.\ref{extract_metric}. Then we have presented the governing equation of the scalar field propagating on this type of oscillating space-time. In Sec.\ref{def_abs}, we have discussed the way we define the absorption cross section in terms of energy flux for an oscillating background like the ringing Schwarzschild black hole, more details on this are also outlined in  Appendix.\ref{cal_abs}. In Sec.\ref{numerics}, we have presented the numerical results of the Absorption cross section. Finally, in Sec.\ref{observ} we have depicted the possibility of observing the effect of oscillating scalar waves on the ringing b\texttt{\textsf{•\textit{•}}}ackground, giving rise to the superradiance.
 
\section{Background and Framework}\label{metric}
For simplicity we consider ringing Schwarzschild BH with metric,
$ 
g_{\mu\nu}\rightarrow{g^s_{\mu\nu}+h_{\mu\nu}}
$,
where $g_{\mu\nu}^s$ is the standard Schwarzschild metric. $h_{\mu\nu}$ is the gravitational wave (GW) perturbation.
We further consider the quadrupole oscillation, $l_0=2$, and azimuthal mode, $m_0=0$ (QNM frequency does not depend on $m_0$ \cite{Regge:1957td}), of the Schwarzschild metric. The oscillating (ringing) metric ( see Appendix.\ref{extract_metric} for details) is  expressed in radiation gauge \cite{Zerilli:1971wd}, which entails the correct asymptotic behavior of the GW flux, 
\be
\bea
&h_{\mu\nu}=\frac 12e^{-\mi\omega{t}}\begin{pmatrix}
	Hf(r)Y_2^0 & H_1Y_2^0 & 0 & 0\\
	H_1Y_2^0 & Hf(r)^{-1}Y_2^0 & h^{(e)}_{1}\pr_\theta{Y_2^0} & h^{(o)}_1s_\theta\pr_\theta{Y_2^0}\\
	0 & h^{(e)}_{1}\pr_\theta{Y_2^0} & r^2\mathcal{T}_2^0 & \frac{1}{2}h_2\mathcal{I}_2^0\\
	0 & h^{(o)}_1s_\theta\pr_\theta{Y_2^0} & \frac{1}{2}h_2\mathcal{I}_2^0 & r^2s^2_\theta\tilde{\mathcal{T}}_2^0
\end{pmatrix} + h.c. 
\eea
\ee
where, symbols are, $\mathcal{I}_2^0=(c_\theta\pr_\theta{Y_2^0}-s_\theta\pr^2_\theta{Y_2^0})$, 
$\mathcal{T}_2^0=K^{(N)}Y_2^0+G^{(N)}\pr^2_\theta{Y_2^0}$, and
$\tilde{\mathcal{T}}_2^0=K Y_2^0+\cot\theta{G} \pr_\theta{Y_2^0}$. $s_\theta=\sin\theta$, $c_\theta=\cos\theta$ and $Y_{lm}$ is the spherical harmonics. The time dependent part of the ringing fluctuation is expressed as $ e^{-i \omega t}$, with $\omega$ being quasi-normal frequency. The perturbation variables are divided into parity odd ($h^{(o)}_1,h_2$) and parity even $(h^{(e)}_1,H, H_1, K,G)$ function which are functions of radial coordinate $r$. 
The Einstein's equation governing the ringing perturbation variables boils down to the well known Regge-Wheeler equations\cite{Regge:1957td,Edelstein:1970sk,Zerilli:1970se}
\be\label{regee-wheeler.eq}
\frac{d^2 \tilde{Z}_i}{dr_*^2}+(\omega^2-V_{i}){\tilde Z}_i=0 ,
\ee
where, $i\equiv ({\cal{E}},{\cal O})$ are assoicated with ${\cal{E}}$ven and ${\cal O}$dd perturbation. $r^* = r + 2 M \ln(r/2M -1)$ is the Tortoise coordinate. For quadrupole oscillation,  the potentials assume the following form,
\begin{eqnarray}
V_{\cal{E}} &=& f(r) \frac{8(3r^3 + 3 Mr^2 )+18{M^2}(2r+M)}{r^3(2{r}+3M)^2} \nonumber\\
V_{\cal{O}} &=& f(r)\Big(\frac{6}{r^2}-\frac{6M}{r^3}\Big) ,
\end{eqnarray}
Where, $f(r) =1-{2M}/{r}$ is the Schwarzchild metric function and. $M$ is mass of the black hole. The functional dependence of odd parity variables on $\tilde{Z}_{odd}(r)$ and even parity variables on $\tilde{Z}_{even}(r)$ are explicitly derived in \cite{Zerilli:1971wd, Zerilli:1970se}. The near horizon values of the ringing fields will be parameterized by $(|\tilde{Z}_{odd}(r\to{2M})|= {\cal O}_h,|\tilde{Z}_{even}(r\to{2M})| = {\cal E}_h)$.

Our goal is to compute the absorption cross-section of the ultra-light scalar field in the ringing Schwarzchild background just described. The equation of a minimally coupled scalar field is  
\be \label{eomphi}
\Box_g \phi+\mu^2 \phi=0 .
\ee
where $\mu$ is the mass of the scalar field.Importantly, in our analysis we can always maintain scalar field amplitude $\phi  < h^\mu_\mu$ by multiplying small number as \eqref{eomphi} remains invariant under constant scaling. Using such scaling, scalar energy momentum tensor can always be made subleading compared to that of the GW background. With this assumption we solve scalar field perturbativly in terms of GW fluctuation, $h_{\mu\nu}$. For simplicity, most of our discussion will be for $\mu \approx 0$. One of the potential candidate scalar fields could be ultra-light axion which is of interest as a potential dark matter\cite{Chadha-Day:2021szb} candidate.  Considering the spherical harmonic expansion of the scalar field, $\phi=\sum {\cal N}_{lm}\xi^{lm}(t,r)Y_{lm}(\theta,\phi)$, where ${\cal N}_{lm}$ being normalization constant, \eqref{eomphi} can be transformed into,
\be\label{scalar1}
{\cal L}_s \xi^{lm} + \sum_{c\gamma} {\cal P}_{lmc\gamma}(h) \xi^{c\gamma} + 
\sum_{c\gamma} \bar{\cal P}_{lmc\gamma}(h^*) \xi^{c\gamma} = 0 .
\ee
Where ${\cal L}_s$ is the radial differential operator corresponding to static Schwarzschild space-time ; 
\be
{\cal L}_s\xi^{lm} =-\pr^2_t\xi^{lm}(t,r)+f(r)\frac{1}{r^2}\pr_r\{r^2f(r)\pr_r\xi^{lm}(t,r)\}\nonumber-f(r)\frac{l(l+1)}{r^2}\xi^{lm}(t,r)
\ee
 And ${\cal P}_{lmc\gamma},\bar{\cal P}_{lmc\gamma}$, (see Appendix.\ref{Source_terms} for detail expression) are the differential operators dependent on the first-order complex and its conjugate part of the metric fluctuation $``h"$ respectively. Gravitational-wave background naturally breaks spherical symmetry, due to which different angular momentum modes of the scalar field will be coupled to each other. Following perturbative approach, the field $\xi^{lm}$ is expanded as,
\be \label{expand}
\xi^{lm}(t,r)=\xi^{lm}_0(t,r)+\alpha\xi^{lm}_{(1)}(t,r)+...,
\ee
where, $\alpha$ counts the order of $h_{\mu\nu}$.
Now, we solve 
\be\label{eom_expand}
\bea
 &{\cal L}_s \xi^{lm}_0 = 0, \\
 &{\cal L}_s\xi^{lm}_{(1)} + \sum_{c\gamma} {\cal P}_{lmc\gamma} \xi^{c\gamma}_0+\sum_{c\gamma} \bar{\cal P}_{lmc\gamma} \xi^{c\gamma}_0= 0 .
\eea
\ee
By using the properties of the inhomogeneous differential equation, we decomposed, $\xi^{lm}_{(1)}= \xi^{lm}_{1}+\bar\xi^{lm}_{1}$, such that,  
\be\label{part.sceq}
{\cal L}_s\xi^{lm}_{1} + \sum_{c\gamma} {\cal P}_{lmc\gamma} \xi^{c\gamma}_0=0~;~{\cal L}_s \bar{\xi}^{lm}_{1}+\sum_{c\gamma} \bar{\cal P}_{lmc\gamma} \xi^{c\gamma}_0= 0
\ee 
The above set of equations can be thought of as scalar waves propagating in the static Schwarzchild ($g_{\mu\nu}^s$) background with oscillatory source term \citep{Siddhartha:2019yjm}. Therefore, to solve this we set the following ingoing initial boundary condition near the horizon of the static black hole,
\be
 \xi_0^{lm}={\cal N}_0  \left(\zeta^{lm}_0 + (r-2M) \frac{(16M^2\km^2-l(l+1))}{2M(4\mi{M}\km-1)}\zeta^{lm}_0 + ...\right) 
 \label{bcond}
\ee
with arbitrary normalization constant $\zeta^{lm}_0$. ${\cal N}_0 = e^{-i \km t} f(r)^{-2 i M \km}$, which corresponds to the ingoing mode near the Schwarzschild horizon of momentum $\km$.  

\section{Defining the absorption cross section}\label{def_abs}
\begin{figure}[t]
{\includegraphics[scale=0.25]{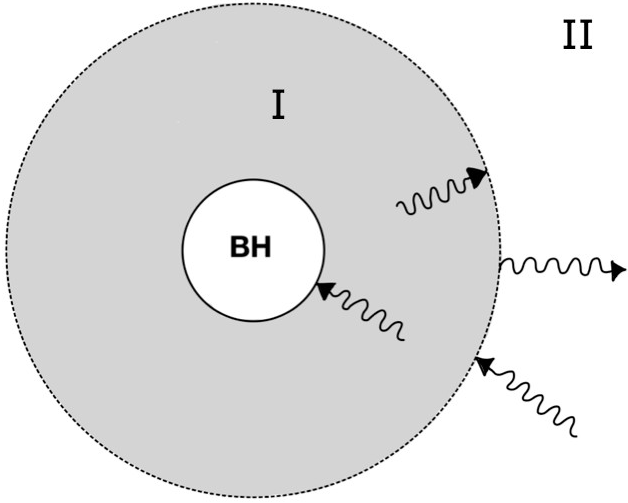}}
\caption{Interaction of the scalar wave with the gravitational wave. The shaded region(I) represents the spatial extent of ringing fluctuation, outside (II) is considered to be static Schwarzschild space-time.}\label{rint}
\end{figure}

Usually one defines the absorption cross-section in the asymptotic flat space region. To compute such quantity for the ringing black hole we propose the following method: Spatial section is divided into region-I (shaded) with ringing background and region-II with pure Schwarzschild (see Fig.\ref{rint}). Between the regions, a hypothetical surface, named as "interaction surface", is defined at $r=r_{int}$, where at $t=t_{int} = r^*_{int}$, the incoming scalar wave interacts with the gravitational wave. 
Considering Eq.\eqref{bcond}, we  numerically solve for each mode of the scalar field up to the first order in $h$ in the region-I described as,
\be\label{tot_sol}
\xi_{\bf I}^{\km{lm}}(t,r)=\xi^{lm}_0 + \xi^{lm}_{1}+\bar\xi^{lm}_{1} +\cdots
\ee
In the region-II, since we consider the pure Schwarzschild background, the solution will be 
\be\label{region2}
\xi_{\bf II}^{\km{lm}}(t,r)=\xi^{lm}_s ~~\mbox{with} ~~{\cal L}_s \xi^{lm}_s = 0 ,
\ee
The above equation is a second order partial differential equation. To solve such Eq.\ref{region2} in Region-II, the natural time-dependent boundary condition can be taken as, 
\be
\bea
\label{bcs}
&\xi^{\km{lm}}_{\bf II}(t,r)|_{\forall{t},r\to{r_{int}}}=\xi_{\bf I}^{\km{lm}}(t,r)|_{\forall{t},r\to{r_{int}}}\\
&\pr_{r}\xi_{\bf II}^{\km{lm}}(t,r)|_{\forall{t},r\to{r_{int}}}=\pr_r\xi_{\bf I}^{\km{lm}}(t,r)|_{\forall{t},r\to{r_{int}}}\\
&\xi_{\bf II}^{\km{lm}}(t,r)|_{t\to{\infty},\forall{r}}={{\xi}}^{{\km}lm}_0(t,r)|_{t\to{\infty},\forall{r}}\\
& \pr_t\xi_{\bf II}^{\km{lm}}(t,r)|_{t\to{\infty},\forall{r}}=-\mi\km{{\xi}}^{{\km}lm}_0(t,r)|_{t\to{\infty},\forall{r}}
\eea
\ee
Within the light cone the quasi-normal oscillation is exponentially decaying in time. Hence, in $t\rightarrow \infty$ fluctuation part of the scalar field $(\xi^{lm}_{1},\bar\xi^{lm}_{1})$ at the interaction surface vanishes. This essentially sets our last two boundary conditions which make sure that the scalar field absorption cross-section of ringing BH boils down to static Schwarzschild value within the characteristic time scale of the oscillation $\tau \sim 2\pi/\omega$.
Using the boundary condition Eq.\ref{bcs}, we solve Eq.\ref{region2} within the region $t\geq{r_*}$, of the box bounded in $(r,t)$ plane as $([r_{int}, \infty], [t_{int}, \infty])$. Asymptotically, the gravitational wave propagates along the outgoing null coordinate. Hence, once solution is obtained, we transform it into outgoing null coordinate  $(t,r)\to(u = t-r^*,r) $ and define the absorption cross-section, which will naturally be $r_{int}$ dependent. 

\subsection{Energy Flux:}
In outgoing null coordinate, the energy flux (see Appendix.\ref{cal_abs} for details) measured by a stationary observer at $r\to\infty$ is, 
\be
\pr_u\mathcal{E}=\int{d}\Omega{r^2}[{\mathcal{T}_{ru}}-{\mathcal{T}_{uu}}]
\ee
and associated absorption cross-section for individual $\km$ mode is defined as 
\be
\sigma^{\km l}_{ring}(u,r_{int})=\frac{\pr_u\mathcal{E}^{\km{l}}}{\pr_u\mathcal{E}_{in}^{\km}},
\ee
where $\mathcal{T}'$s are the components of the energy-momentum tensor. $\pr_u\mathcal{E}_{in}^{\km}$ is the energy flux of the incoming plane wave defined at the infinity. The methodology we adopted to solve time dependent system is not particularly suitable to identify the appropriate normalization. To proceed, we note the approximate asymptotic solution as,  
\be\label{asymp}
\xi^{{\km}lm}_s(t,r)=\frac{\mathcal{A}^{\km}_{{l}m}(u,r_{int})}{r}e^{-\mi\km(u+2r_*)}+\frac{\mathcal{B}^{\km}_{{l}m}(u,r_{int})}{r}e^{-\mi\km{u}}.
\ee
The ingoing/outgoing coefficients $(\mathcal{A}^{\km,\omega}_{{l}m}/\mathcal{B}^{\km,\omega}_{{l}m})$ will match with that of the static Schwarzschild case in $u\rightarrow \infty$ limit. Following the standard procedure \cite{Unruh:1976fm},  assumption of incoming wave along z-direction in the asymptotic infinity sets the constant normalization factor (see Appendix.\ref{cal_abs} for details) to be, $\mathcal{N}_{\km{lm}}=\sqrt{4\pi(2l+1)}\delta_m^0/(2\mi\km{\mathcal{A}^{\km}_{{l}m}(u\to\infty,r_{int}))}$. Using this, the ingoing plane wave at the asymptotic infinity is assumed as,
\be
\phi_{in}\sim\frac{\int\sum \mathcal{N}_{\km{lm}}\mathcal{A}^{\km}_{{l}m}(u,r_{int})Y_{lm}(\Omega){d\Omega}}{\int\sum\sqrt{4\pi(2l+1)}Y_l^0(\Omega){d\Omega}}2\mi{\km}e^{-\mi\km(u+2z_*)}.
\ee
Where, time dependent amplitude is averaged over angle.
 With all these ingredients we will now numerically compute the absorption cross-section $\sigma^{\km l}_{ring}$. 

\section{Numerical Computation of $\sigma^{\km l}_{ring}$}\label{numerics}
As mentioned earlier, the gravitational wave background is assumed to be quadrupole oscillation (Eq.\ref{regee-wheeler.eq}) with the QNMfrequency $\omega = (0.74734-\mi ~0.17792)(r_h)^{-1} $ \cite{Chandrasekhar:1975zzb}  , where,  Schwarzschild horizon radius $r_h=2M$.
We express all physical parameters in unit of $r_h$.
Anyway, the frequency is known to be the same for both even and odd parity perturbation \cite{Chandrasekhar:1975zza,Chandrasekhar:1975zzb,Fiziev:2019ewy}. Ringdwon phase in general should contain all possible QNM modes\cite{Chandrasekhar:1975zza}. However, we consider the one which is long-lived.
The background ringing field solutions are so chosen that the perturbativeness defined as $\delta g/g_s \propto h^\mu_\mu \ll 1 $ is maintained for a wide range of initial parameters. 
Given the ringing black hole background with a specific QNM frequency, we solve for the scalar field Eq.\ref{scalar1}. Importantly, we should reproduce the well known static value of the absorption cross-section\cite{Huang:2014nka} associated with Schwarzschild black hole in the limit,  $\lim_{u\rightarrow \infty}\sigma^{\km l}_{ring}(u,r_{int}) = \sigma^s(\km,l)$. Hence, before the static limit is reached over the time scale $\tau \sim 40$, $\sigma^{\km l}_{ring}(u,r_{int})$ will also undergo a ring down phase. It is during the ring-down phase, when the superradiance is observed.

\begin{figure}[t]
{\includegraphics[scale=0.4]{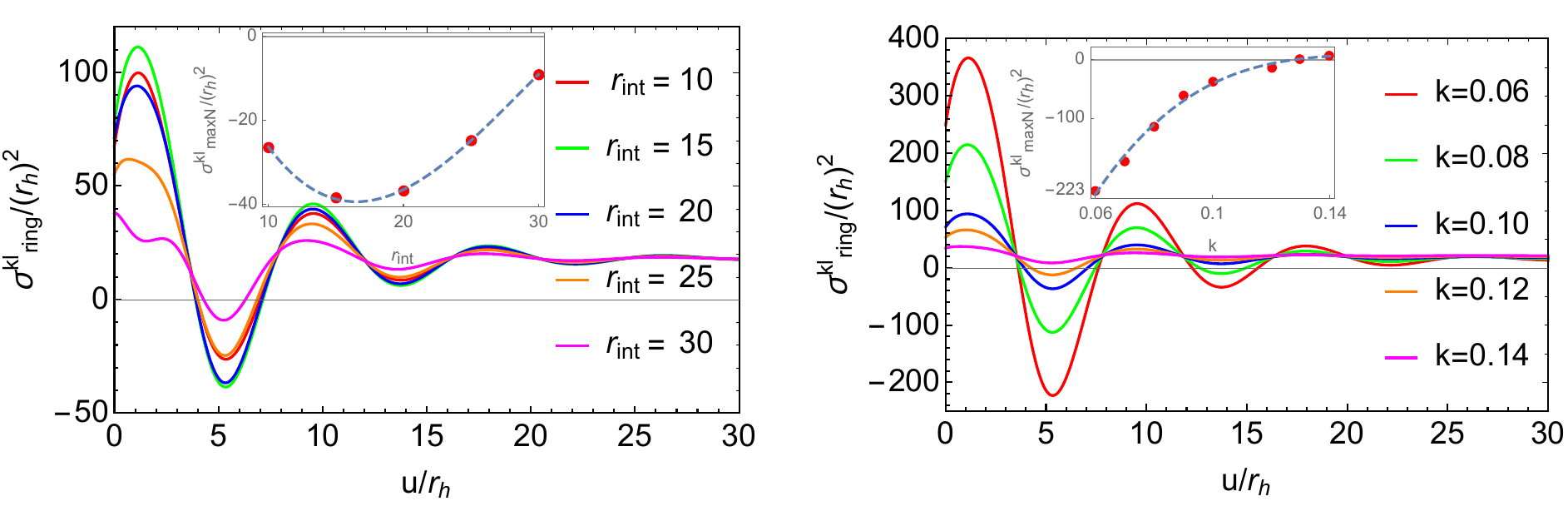}}
\caption{{\bf Left panel}: we have plotted $\sigma^{\km l}_{ring}$ with respect to time considering $l=0,k=0.1$ for different $r_{int}$. In the inset the variation of {\bf max}imum {\bf N}egative value symbolized as $\sigma^{kl}_{\bf maxN}$ is plotted with respect to $r_{int}$
	{\bf Right panel}: we have plotted the same for $l=0$ and vary $\km$. The inset shows the variation of maximum negative value $\sigma^{kl}_{\bf maxN}(k)$ with respect to $\km$. All plots are from $\mu=0$. }\label{varrint_vark}
\end{figure}

\begin{figure}[t]
{\includegraphics[scale=0.4]{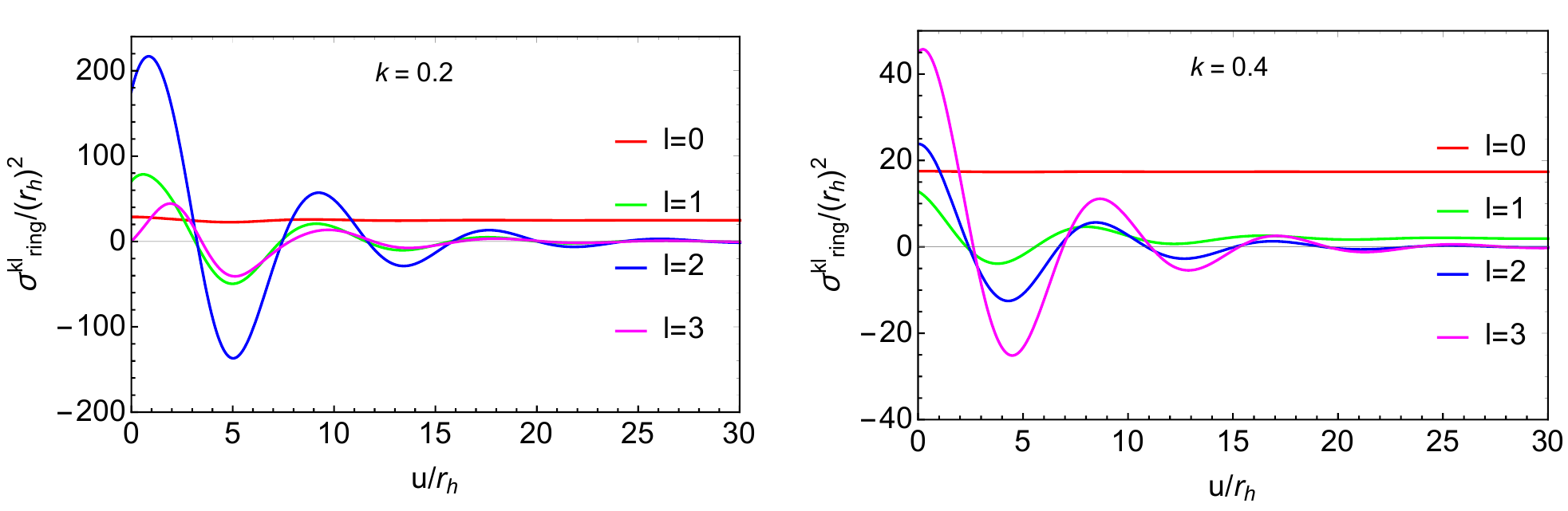}}
\caption{The partial absorption cross-section is plotted with respect to time for two frequencies $k=0.2$(left) and $k=0.4$(right) for different $l$. All plots are from $\mu=0$.}\label{ldiff}
\end{figure}

\begin{figure}[t]
{\includegraphics[scale=0.4]{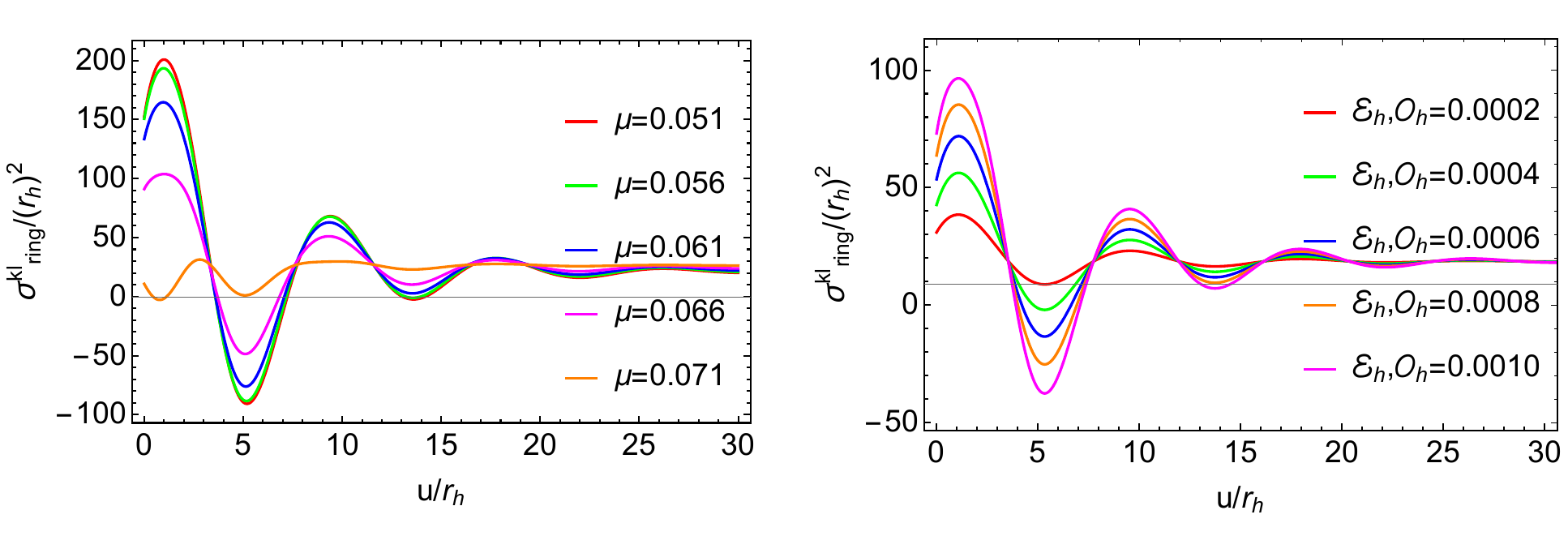}}
\caption{{\bf Left panel}:we have plotted  $\sigma^{\km l}_{ring}$ with respect to time for $l=0,\sqrt{k^2+\mu^2}=0.1,r_{int}=20$ for different mass of the scalar field. {\bf Right panel}: we have plotted the same  for $l=0,k=0.1,r_{int}=20$ and vary ${\cal E}_h ={\cal O}_h$ for massless case, $\mu=0$.}\label{bamp_mass}
\end{figure}

Elaborating more on the numerics, our final solutions have been observed to be stable for a wide range of initial conditions parametrized by $\zeta^{lm}_0$ within $\sim{10^{-2}-20}$. 
Up to a small fluctuation our results are also stable for a range of asymptotic radial infinity within $r = 75r_h - 100 r_h$.   
This fluctuation may be arising due to our approximate normalization. 
Nonetheless, the characteristic features of absorption cross-section for different angular momentum modes have been observed to be the same. Hence, we particularly focus on $l=0$ mode. As emphasized in the begining the most important characteristics emerged out from our study is the super-radiant(negative) absorption cross-section in its ringing phase for all angular momentum mode (see Fig.\ref{varrint_vark}, \ref{ldiff}, \ref{bamp_mass}).

For the given $\omega$ and $(\mathcal{E}_h,\mathcal{O}_h)$, the maximum super-radiant absorption cross-section amplitude symbolized as $\sigma^{kl}_{\bf maxN}(\km)$, decreases with increasing momentum $\km$. We also derived a fitting formula  
$\sigma^{k l=0}_{\bf maxN}(\km)=-1021 + 20065 \km - 131798 \km^2 + 292061 \km^3$, withing the range $(k=0.06\to{0.14})$.
Following our expectation, we observe the existence of a maximum value of $\km_{max}$ above which super-radiance vanishes. However, absorption cross-section will still remains oscillatory with a positive magnitude, and attains its static Schwarzschild value in ring down time scale (see fig.\ref{varrint_vark}).
The physical reason behind vanishing of superradiance can be attributed to the decoupling of higher momentum modes from the gravitational wave fluctuations. Our numerical analysis provides: $l=0,\km_{max}\sim{0.13}$ ; $l=1, \km_{max}\sim{0.45}$ ; $l={2}, \km_{max}\sim{1.0}$ ; $l={3}, \km_{max}\sim{1.5}$ ; $l={4}, \km_{max}\sim{1.8}$ for (${\cal E}_h \simeq {\cal O}_h \sim 10^{-3}$ and $r_{int}=20$). Of course decreasing the background amplitude would make $\km_{max}$ lower.

As our methodology suggests, the ringing phase of the absorption cross-section and its amplitude depend on the location of the interaction surface $r_{int}$ shown in fig.(\ref{varrint_vark}). For each $(l,\km)$ value, there exists a maximum possible super-radiant amplitude ($\sigma^{kl}_{\bf maxN}(r_{int})$) as one varies  $r_{int}$. For example, for $l=0$ it occurs approximately at  $r_{int}\sim{20}$, for higher range $\km$ values.  In the low $\km$ region this location of maximum super-radiant amplitude shift towards higher $r_{int} \sim{30-40}$. This behavior can again be fitted as $\sigma^{k l=0}_{\bf maxN}(r_{int})=63.7 - 14.4 r_{int} + 0.6 r^2_{int}- 0.007 r^3_{int}$ withing the range $(r_{int}=10\to{30})$. Each super-radiant mode has been observed to be saturated to a particular negative value of the absorption cross-section for large $r_{int}$. 
So far we discussed about the absorption cross-section for fixed value of ${\cal E}_h \simeq {\cal O}_h \sim 10^{-3}$. 
However, background gravitational wave amplitude plays a crucial role in enhancing the outgoing amplitude of the scalar wave compared to the incoming one. This fact motivates us to look into the variation of $\sigma^{\km l}_{ring}$ with respect to $({\cal E}_h, {\cal O}_h)$ as shown fig.(\ref{bamp_mass}). Decreasing  background amplitude of $({\cal E}_h, {\cal O}_h)$ reduces the overall amplitude of $\sigma^{\km l}_{ring}$ in its ringing phase as shown in the Fig.(\ref{bamp_mass}), and finally super-radiance ceases to exist at around ${\cal E}_h,{\cal O}_h=10^{-4}$(in units of black hole mass) for $l=0$. This conclusion has been observed to be true for higher $l$ mode as well.
Thus far we discussed a particular angular momentum mode $l=0$. Behavior of $\sigma^{\km l}_{ring}(u,r_{int})$ for different $l$ is important. For a given location of the interaction surface ($r_{int}=20$), the left panel of Fig.\ref{ldiff} shows that in the lower momentum region (k=0.2) the super-radiant amplitude first increases up to $l=2$ and then become suppressed after $l=3$. This does not hold true for all the momentum mode as can be seen in the right panel of Fig.\ref{ldiff} for $k=0.4$. 
Because of non-trivial dependence on $r_{int}$ described before, maximum super-radiant amplitude happens to be at different location of the interaction surface $r_{int}$ for different $(l,k)$. Hence, overall our study suggests that with increasing $l$ the enhancement of super-radiance amplitude can be attributed to  ``mode-mixing" and an increasing number of modes contributing as a background source term in Eq.\ref{eom_expand}. A similar kind of feature has been observed for moving black holes where absorption cross-section has been shown to diverge logarithmically with angular momentum $l$ \cite{Cardoso:2019dte}.


Finally, we perform preliminary analysis for  (fig.(\ref{bamp_mass})) massive scalar. What we observed is that for a given mode, $(\tilde{\omega}=\sqrt{k^2+\mu^2}, l)$ and $r_{int}$, as we increase the mass of the scalar field, supperradiant amplitude increases towards a maximum value and then after it decreases towards zero for a critical value of $\mu_{cri} < \tilde{\omega}$. For example $\mu_{cri}\sim{0.071}$ in (fig.\ref{bamp_mass}) for $l=0,\tilde{\omega}=0.1,rint=20$. Our primary observation is that along with the increasing $l$, the $\mu_{cri}$ is increasing approximately linearly. Detailed study for the massive scalar and it bound will be reported elsewhere.

\section{Observables effects of the oscillating scalar field}\label{observ}
\begin{figure}[t]
{\includegraphics[scale=0.4]{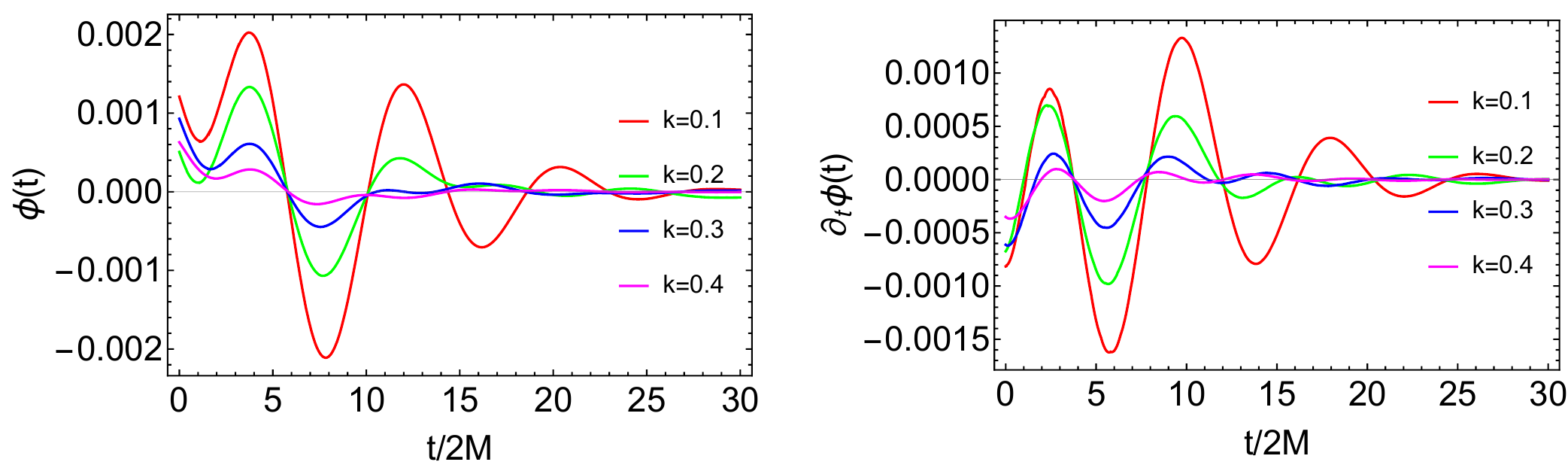}}
\caption{Real part of the outgoing axion wave, subtracting the contribution from static black hole, has been plotted in the left panel, with respect to time for a fixed frequency, $\km=(0.1, 0.2, 0.3, 0.4)$. And the time derivative of the same quantity has been plotted in the right panel.Time is measured from a point $t_{inf}=r_{inf}$ on the light cone. The location of the interaction surface is taken as $r_{int}=30$}\label{probrotedm}
\end{figure}

Superradiance phenomena is known to occur in the context of static Kerr and charged black hole \cite{Benone:2019all}. However, most striking feature of our present study is its oscillatory nature which is observed to carry the information of black hole though their quasi-normal modes. 
%
Any time varying observable is always physically motivating when it comes to observation compared to the static one. Identifying the ringing scalar field as axion, we calculate multiple observables, which can in principle be observed in laboratory \cite{Irastorza:2021tdu,Schumann:2019eaa}. Treating axion field as time varying background, and considering well known interaction term, $\mathcal{L}_{int}\sim g \phi F_{\mu\nu}\tilde{F}^{\mu\nu}$,
	we calculated time varying induced rotation of linear photon polarization  expressed as $\theta=-\mi(\tilde{\Delta}_\phi-\tilde{\Delta}^*_\phi)\sim g~\phi(t)$ (see Appendix.\ref{photon_axion}). Where,  $\tilde{\Delta}_\phi=\int^t\frac{\mi}{2} g ~\pr_{t'} {\phi}$, with axion-photon coupling taken as, $g\sim10^{-13}-10^{-14} ~\mbox{GeV}^{-1}$ \cite{Fedderke:2019ajk}. 
Experimental searches of such rotation due to background axion has been extensively studied in the literature \cite{Obata:2018vvr, Fujita:2018zaj}. Our analysis suggests, there would be an extra time varying contribution originating from the ringing oscillation, that can in principle be observed in near future. 	
	The time varying nucleon electric dipole moment $(N_{edm})$, is calculated as $N_{edm} = h\, \phi(t)$, considering the following nucleon(N)-axion-photon interaction $\mathcal{L}\sim -\frac{\mi}{2} h \phi \bar{N}\sigma_{\mu\nu}\gamma_5 NF^{\mu\nu}$. The value of axion-nucleon coupling  $h$, is typically set from decay constant for QCD axion \cite{Graham:2013gfa}. The time profile of the above mentioned two observables are depicted in the left panel of Fig.\ref{probrotedm} for different momentum of the axion.

Finally we consider axial axion-fermion type coupling $\mathcal{L}\sim -\frac{\mi}{2} \zeta \partial_{\mu} \phi \bar{\psi}\gamma^{\mu}\gamma^{5} \psi$, which describes a physical system where spin of the fermion will be precessing around the direction of local momentum, ${\bf v}\,\pr_t \phi$, of the axion, as can be seen from the Hamiltonian arising due to this coupling, $H \sim \zeta \pr_t\phi\, {\bf v}\cdot\pmb{\sigma}_\psi$, where $\pmb{\sigma}_\psi$ is the fermion spin operator. That leads to a shift in the energy levels of the fermions (nuclear or electron), $\Delta E_{nm}(t)\sim \zeta  |{\bf v}| \pr_t\phi$ due to its axial moment. The coupling parameter $\zeta\sim 10^{-9}~ \mbox{GeV}^{-1}$, \cite{Raffelt:2006cw} is constrained from supernova cooling rates, and ${\bf v}$ is the relative velocity (in astrophysical context galactic virial velocity $|{\bf v}|\sim 10^{-3}$ may be used) between axion and fermion \cite{Graham:2013gfa}. The time variation of this energy shift is shown in right panel of the Fig.\ref{probrotedm}. Detecting such extra time varying contribution to the energy shift is promising given the  several existing proposal of measuring those quantities \cite{Graham:2011qk} using the method of ``Precision magnetometry" using cold molecules \cite{Budker:2013hfa}.
	In order to show oscillating features of all the observables, we consider s-wave($l=0$) outgoing mode of the ringing axion and subtracted the effect due to static black hole.  Time is measured from a point $t_{inf}=r_{inf}$ on the light cone, where the detector is assumed to be placed \cite{Nagano:2021kwx, Chigusa:2019rra}. 
\section{Conclusions}
In spite of being widely discussed in the literature, recent observation of gravitational waves has led to a resurgence of exploring the phenomena of black hole superradiance in a more general gravitational setting. Such phenomena were so far shown to exist in Kerr and charged black hole background. In this paper, we report for the first time that the black hole in its ringing phase can also lead to superradiance when interacting with an incoming scalar wave. Apart from detecting gravitational waves, our present study opens up interesting possibilities of observing black hole merging phenomena through the complementary observables in terms of other fundamental fields. The basic mechanism behind this superradiance is simple. When an incoming scalar wave passes through the ringing GW background, gravitational energy can be transferred into the scalar wave leading to the enhancement of its out-going flux. This is precisely what makes the scalar field absorption cross-section $\sigma^{\km l}_{ring}$ also going through the ringing phase. Before settling down to its standard Schwarzschild value, $\sigma^{\km l}_{ring}$ in its ringing phase assumes the negative value indicating the superradiance phenomena in the ringing black hole background. Finally we computed different possible time varying observables through axion-photon, axion-fermion coupling. All the observables namely, rotation of photon polarization ($\theta(t)$), nucleon electric dipole moment ($N_{edm}(t)$) and shifting of the energy levels ($\Delta E_{nm}(t)$) due to fermionic axial moment, naturally encode the ringing oscillation through the axion.
  
{\bf Acknowledgement: }We would like to thank our Gravity and High Energy Physics groups at IIT Guwahati for illuminating discussions.
We also thank the anonymous referee for their valuable comments.
\appendix
\section{ Metric corresponding to black hole ring down phase in radiation gauge}\label{extract_metric}
In this section we will discuss how the metric describing the ringing black hole came about, mainly following two defining papers \cite{Regge:1957td}\cite{Zerilli:1971wd} (for corrected typos one may refer to \cite{Edelstein:1970sk}) of black hole perturbation theory. We consider gravitational perturbation, $h_{\mu\nu}$, on a static Schwarzschild metric ($g^s_{\mu\nu}$) as,
\be 
g_{\mu\nu}\rightarrow{g^s_{\mu\nu}+h_{\mu\nu}}
\ee
Regee-Wheeler\cite{Regge:1957td} found that the perturbation, $h_{\mu\nu}$, can be decomposed into even and odd parity and it was shown  that the odd parity perturbation equations would boil down to a Schrodinger like second order differential equation. Zerrili\cite{Zerilli:1971wd} found the same type of equation for even parity perturbation. Here We will briefly review the gauge choice given in Zerilli, keeping up the notations (\emph{supressing the space-time index, $\mu,\nu$, all the tensorial quantity written in bold format using spherical harmonics index, $l,m$, of the tensor perturbation}) intact. Putting together, the Even ($``e"$) and odd parity ($``o"$) perturbation can be written as
\be
{\bf h}_{lm}=\sum_{lm}[\bf{h}^{(e)}_{lm}+\bf{h}^{(o)}_{lm}]
\ee
where,
\be
{\bf h}^{(o)}_{lm}=\frac{\mi}{r}\sqrt{2l(l+1)}\Big[\mi{h_{0lm}}(t,r){\bf c}^{(0)}_{lm}+{h_{1lm}}(t,r){\bf c}_{lm}-\frac{1}{2r}\sqrt{(l-1)(l+2)}h_{2lm}(t,r){\bf d}_{lm}\Big]
\ee
and
\be
\bea
{\bf h}^{(e)}_{lm}&=\Big(1-\frac{2M}{r}\Big){H_{0lm}}(t,r){\bf a}^{(0)}_{lm}-\sqrt{2}\mi{H_{1lm}}(t,r){\bf a}^{(1)}_{lm}+\Big(1-\frac{2M}{r}\Big)^{-1}{H_{2lm}}(t,r){\bf a}_{lm}\\
&-\frac{1}{r}\sqrt{2l(l+1)}\Big(\mi{h^{(e)}_{0lm}}(t,r){\bf b}^{(0)}_{lm}+\mi{h^{(e)}_{1lm}}(t,r){\bf b}_{lm}\Big)\\
&+\sqrt{\{l(l+1)(l-1)(l+2)/2\}}{G_{lm}}(t,r){\bf f}_{lm}+\sqrt{2}\{K_{lm}(t,r)-\frac{1}{2}G_{lm}(t,r)\}{\bf g}_{lm}
\eea
\ee
The explicit form of the basis (${\bf a}^{(0)}_{lm}...{\bf g}_{lm}$) can be found in \cite{Zerilli:1971wd}. However, for completeness let us provide the expressions of those ten independent basis tensors in terms of ${\cal G}^{\alpha\beta}_{\mu\nu} = (\delta^{\alpha}_{\mu} \delta^{\beta}_{\nu} +\delta^{\beta}_{\mu} \delta^{\alpha}_{\nu})$ as, 
\be
\bea
&({\bf a}^{(0)}_{lm})_{\mu\nu}= \frac 1 2 Y_{lm} {\cal G}^{tt}_{\mu\nu}   ~~,~~
({\bf a}^{(1)}_{lm})_{\mu\nu}=\frac{\mi}{\sqrt{2}}  Y_{lm} {\cal G}^{tr}_{\mu\nu} ~~,~~
({\bf a}_{lm})_{\mu\nu}=\frac 1 2 Y_{lm} {\cal G}^{rr}_{\mu\nu}~ ,~\\
&({\bf b}^{(0)}_{lm})_{\mu\nu}=\frac{\mi r}{\sqrt{2l(l+1)}}(\pr_\theta Y_{lm} {\cal G}^{t\theta}_{\mu\nu} +\pr_\phi Y_{lm} {\cal G}^{t\phi}_{\mu\nu} ) ~~,~~
({\bf b}_{lm})_{\mu\nu}=\frac{r}{\sqrt{2l(l+1)}} (\pr_\theta Y_{lm} {\cal G}^{r\theta}_{\mu\nu} +\pr_\phi Y_{lm} {\cal G}^{r\phi}_{\mu\nu} )~,~ \\
&({\bf c}^{(0)}_{lm})_{\mu\nu}=\frac{r}{\sqrt{2l(l+1)}}
\left(\frac{1}{s_\theta}\pr_\phi Y_{lm} {\cal G}^{t\theta}_{\mu\nu}- s_\theta\pr_\theta Y_{lm} {\cal G}^{t\phi}_{\mu\nu} \right) ~,~\\
&~({\bf c}_{lm})_{\mu\nu}=\frac{\mi r}{\sqrt{2l(l+1)}} 
\left(\frac{1}{s_\theta}\pr_\phi Y_{lm} {\cal G}^{r\theta}_{\mu\nu}- s_\theta\pr_\theta Y_{lm} {\cal G}^{r\phi}_{\mu\nu}\right)~,~\\
&({\bf d}_{lm})_{\mu\nu}=\frac{-\mi r^2}{\sqrt{2l(l+1)(l-1)(l+2)}}
\left(-\frac{1}{2 s_\theta} X_{lm}{\cal G}^{\theta \theta}_{\mu\nu} + s_\theta W_{lm} {\cal G}^{\theta \phi}_{\mu\nu} + \frac {s_\theta}{2} X_{lm} {\cal G}^{\phi\phi}_{\mu\nu}\right) ~,~\\
&({\bf g}_{lm})_{\mu\nu}=\frac{r^2}{2\sqrt{2}}
(Y_{lm}{\cal G}^{\theta \theta}_{\mu\nu} + s_\theta^2 Y_{lm} {\cal G}^{\phi\phi}_{\mu\nu} )~,~\\
&({\bf f}_{lm})_{\mu\nu}=\frac{r^2}{\sqrt{2l(l+1)(l-1)(l+2)}}
\left(\frac{1}{2} W_{lm}{\cal G}^{\theta \theta}_{\mu\nu} + X_{lm} {\cal G}^{\theta \phi}_{\mu\nu} - \frac {s_\theta^2}{2} W_{lm} {\cal G}^{\phi\phi}_{\mu\nu} \right)~,~
%
\eea
\ee
where, $X_{lm}=2\pr_\phi(\pr_\theta -\cot\theta)Y_{lm}$ and $W_{lm}=(\pr^2_\theta-\cot\theta\pr_\theta-(1/\sin^2\theta))Y_{lm}$, also. Now there are 10 components (3 for odd parity and 7 for even parity) according to the symmetric condition of the gravitational metric. 
Under diffeomorphism, $x_\mu\to{x_\mu+\zeta_\mu}$, the perturbation ${\bf h}_{lm}$ transforms as
\be\label{gauge.trans}
\tilde{{\bf h}}_{lm}={\bf h}_{lm}-2[\nabla\zeta_{lm}]_{s}
\ee
where, the symmetric covariant derivative acts as, $[\nabla\zeta_{lm}]_{s}\to (\nabla_\mu\zeta^{lm}_\nu+\nabla_\nu\zeta^{lm}_\mu$)/2. So we can fix total 4 components out of the 10 components of the perturbation.

For the {\bf odd parity} part of the perturbation we consider the gauge transformation as, 
\be
\zeta^{(o)}_{lm}=\frac{\mi}{r}\Lambda_{lm}(t,r)(0,0, {\bf L}Y_{lm}(\Omega))
\ee
where, ${\bf L}=-\mi\{\hat{e}_\phi\pr_\theta-\hat{e}_\theta(1/sin\theta)\pr_\phi\}$, $\hat{e}_\theta$ and $\hat{e}_\phi$ are the unit vectors along $\theta$ and $\phi$ respectively. Consequently the components of the perturbation transform \eqref{gauge.trans} as 
\be
\bea
{\bf h}^{(o)}_{lm}&=\frac{\mi}{r}\sqrt{2l(l+1)}\Big[\mi\Big\{{h_{0lm}}(t,r)-\frac{\pr \Lambda_{lm}}{\pr t}\Big\}{\bf c}^{(0)}_{lm}(\Omega)+\Big\{{h_{1lm}}(t,r)-r^2\frac{\pr}{\pr r}\Big(\frac{\Lambda_{lm}}{r^2}\Big)\Big\}{\bf c}_{lm}(\Omega)\\
&-\frac{1}{2r}\sqrt{(l-1)(l+2)}(h_{2lm}(t,r)+2\Lambda_{lm}(t,r)){\bf d}_{lm}(\Omega)\Big]
\eea
\ee
One of the components can be fixed to be zero by adjusting $\Lambda_{lm}(t,r)$. To get the suitable asymptotic behaviour of the gravitational wave flux the asymptotic nature of the components should be fixed in such a way that they behave as $1/r$ \cite{price1969} near spatial infinity. For which we resort to fix  $h^{(N)}_{0lm}(t,r)$ to be zero, 
\be
 h^{(N)}_{0lm}(t,r)=0\implies{h_{0lm}}(t,r)=\frac{\pr \Lambda_{lm}}{\pr t}\implies \Lambda_{lm}=-\frac{1}{\mi\omega}h_{0lm}(t,r)
\ee
 the temporal part of the perturbation has been taken as $e^{-\mi\omega{t}}$. \emph{At this moment we want to state that in the following discussion $``N"$ signifies the transformed component and the quantities without $``N"$  denotes the components in Regee-Wheeler's \cite{Regge:1957td} gauge}. The motivation for the modification of gauge transformation stems from the fact that one would get the suitable asymptotic behaviour of the gravitational wave flux in $r\to\infty$. Now coming back, because of the new gauge we get the components as 
\be
 h^{(N)}_{2lm}(t,r)=h_{2lm}(t,r)+2\Lambda_{lm}(t,r)=2\Lambda_{lm}(t,r)=-\frac{2}{\mi\omega}h_{0lm}(t,r)
\ee
(for the functional form of the perturbation parameters written in Regee-Wheeler gauge, look at \cite{Zerilli:1971wd}) and another non zero component
\be
h^{(N)}_{1lm}(t,r)={h_{1lm}}(t,r)-r^2\frac{\pr}{\pr r}\Big(\frac{\Lambda_{lm}}{r^2}\Big) 
\ee
Einstein equation\cite{Edelstein:1970sk} in Regge-Wheeler gauge, governing the odd parity perturbation gives
\be
\omega^2h_{1lm}-\frac{\mi\omega{dh_{0lm}}}{dr}+\frac{2\mi\omega{h_{0lm}}}{r}-(r-2M)(l-1)(l+2)\frac{h_{1lm}}{r^3}=0
\ee
Substituting $h_{0lm}(t,r)=-\mi\omega\Lambda_{lm}(t,r)$,
\be
\omega^2h_{1lm}-\omega^2\frac{d\Lambda_{lm}}{dr}+\omega^2\frac{2}{r}\Lambda_{lm}-(r-2M)(l-1)(l+2)\frac{h_{1lm}}{r^3}=0
\ee
using $\lambda=(l-1)(l+2)/2$, we get from this equation 
\be
h^{(N)}_{1lm}(t,r)=\frac{2\lambda}{\omega^2r^2}\Big(1-\frac{2M}{r}\Big)h_{1lm}
\ee
Considering only the quadrupole perturbation ($l=2$) and choosing $m=0$, we get the metric corresponding to odd parity perturbation in Zerilli gauge as,
\be
h^{Odd}_{\mu\nu}=\begin{pmatrix}
0 & 0 & 0 & 0 \\
0 & 0 & 0 & h^{(N)}_1(t,r) s_\theta\pr_\theta{Y_2^0}\\
0 & 0 & 0 &  \frac{1}{2}h^{(N)}_2(t,r)(c_\theta\pr_\theta{Y_2^0}-s_\theta\pr^2_\theta{Y_2^0})\\
0 & h^{(N)}_1(t,r) s_\theta\pr_\theta{Y_2^0} & \frac{1}{2}h^{(N)}_2(t,r)(c_\theta\pr_\theta{Y_2^0}-s_\theta\pr^2_\theta{Y_2^0}) & 0
\end{pmatrix}
\ee

For {\bf Even Parity} part of the perturbation we consider the following gauge transformation, 
\be
\zeta^{(e)}_{lm}=\mcM_0(t,r)Y_{lm}(\Omega)\hat{e}_t+\mcM_1(t,r)Y_{lm}(\Omega)\hat{e}_r+\mcM_2(0, \nabla{Y_{lm}})
\ee
where, ${\bf \nabla}=\hat{e}_\theta\pr_\theta+\hat{e}_\phi(1/sin\theta)\pr_\phi$, consequently the even parity components of the perturbation transform \eqref{gauge.trans} as 
\be\label{even_gauge}
\bea
{\bf h}^{(e)}_{lm}&=\Big(1-\frac{2M}{r}\Big)\Big\{{H_{0lm}}(t,r)-2\Big(1-\frac{2M}{r}\Big)^{-1}\Big(\frac{\pr \mcM_0}{\pr t}-\frac{M}{r^3}(r-2M)\mcM_1\Big)\Big\}{\bf a}^{(0)}_{lm}\\
&-\sqrt{2}\mi\Big\{{H_{1lm}}(t,r)-\Big(\frac{\pr \mcM_1}{\pr t}+\frac{\pr \mcM_0}{\pr r}-\frac{2M}{r(r-2M)}\mcM_0\Big)\Big\}{\bf a}^{(1)}_{lm}\\
&+\Big(1-\frac{2M}{r}\Big)^{-1}\Big\{{H_{2lm}}(t,r)-2\Big(1-\frac{2M}{r}\Big)\Big(\frac{\pr \mcM_1}{\pr r}+\frac{M}{r(r-2M)}\mcM_1\Big)\Big\}{\bf a}_{lm}\\
&-\frac{\mi}{r}\sqrt{2l(l+1)}\Big\{{h^{(e)}_{0lm}}(t,r)-\Big(\frac{\pr \mcM_2}{\pr t}+\mcM_0\Big)\Big\}{\bf b}^{(0)}_{lm}\\
&+\frac{1}{r}\sqrt{2l(l+1)}\Big\{{h^{(e)}_{1lm}}(t,r)-\Big(\frac{\pr \mcM_2}{\pr r}-\frac{2}{r}\mcM_2+\mcM_1\Big)\Big\}{\bf b}_{lm}\\
&+\sqrt{\{l(l+1)(l-1)(l+2)/2\}}\Big\{{G_{lm}}(t,r)-\frac{2}{r^2}\mcM_2\Big\}{\bf f}_{lm}\\
&+\Big[\sqrt{2}\Big\{K_{lm}(t,r)-\frac{2}{r^2}(r-2M)\mcM_1\Big\}-\frac{l(l+1)}{2}\Big\{G_{lm}(t,r)-\frac{2}{r^2}\mcM_2\Big\}\Big]{\bf g}_{lm}
\eea
\ee
So we can fix three components of the even-parity-components of perturbation by adjusting $\mcM_0,\mcM_1$ and $\mcM_2$. Like we said in the odd parity case, here also we choose the gauge in the following manner. We have found that for the following choice 
\be\label{M1}
\mcM_1(t,r)=\frac{1}{2}r\Big(1-\frac{2M}{r}\Big)^{-1}\Big[\mi\omega-\frac{3M}{\lambda{r^2}}\Big]Z(t,r)
\ee
the perturbation parameter
\be
K^{(N)}_{lm}(t,r)=K_{lm}(t,r)-\frac{2}{r^2}(r-2M)\mcM_1
\ee
(for the functional form of the perturbation parameters written in Regee-Wheeler gauge, look at \cite{Zerilli:1971wd}) goes as $1/r$ in $r\to \infty$, excluding the plane wave part $e^{-\mi\omega(t-r_*)}$ ( as $Z\sim e^{\mi\omega r_*}$).
Form the Einstein equation, governing the perturbation, one can find that     having no source term, $H_0=H_2=H$, accordingly we also fix $H^{(N)}_0=H^{(N)}_2=H^{(N)}$, which implies 
\be\label{M0M1}
\bea
&\Big(1-\frac{2M}{r}\Big)^{-1}\Big(\frac{\pr \mcM_0}{\pr t}-\frac{M}{r^3}(r-2M)\mcM_1\Big)=\Big(1-\frac{2M}{r}\Big)\Big(\frac{\pr \mcM_1}{\pr r}+\frac{M}{r(r-2M)}\mcM_1\Big)\\
&{\implies}\frac{\pr \mcM_0}{\pr t}=\Big(1-\frac{2M}{r}\Big)^2\frac{\pr \mcM_1}{\pr r}+\Big(1-\frac{2M}{r}\Big)\frac{2M}{r^2}\mcM_1\\
&\implies\mcM_0=-\frac{1}{\mi\omega}\Big\{\Big(1-\frac{2M}{r}\Big)^2\frac{\pr \mcM_1}{\pr r}+\Big(1-\frac{2M}{r}\Big)\frac{2M}{r^2}\mcM_1\Big\}
\eea
\ee
( remember time dependence of the perturbation assumed to be $e^{-\mi\omega{t}}$) which fixes $\mcM_0$ in terms of $\mcM_1$.
Making $h^{(e)(N)}_{0lm}=0$ would lead \eqref{even_gauge} to
\be
{\implies}\Big(\frac{\pr \mcM_2}{\pr t}+\mcM_0\Big)={h^{(e)}_{0lm}}(t,r)
\ee
as we had ${h^{(e)}_{0lm}}(t,r)=0$ in Regee-Wheeler gauge, so
\be\label{M0M2}
\bea
&\frac{\pr \mcM_2}{\pr t}=-\mcM_0\\
&{\implies}\mcM_2=\frac{1}{\mi\omega}\mcM_0
\eea
\ee
Substituting $\mcM_0$ in terms of $\mcM_1$\eqref{M0M1} in \eqref{M0M2} $\mcM_2$ gets fixed as
\be
\mcM_2=\frac{1}{\mi\omega}\Big[-\frac{1}{\mi\omega}\Big\{\Big(1-\frac{2M}{r}\Big)^2\frac{\pr \mcM_1}{\pr r}+\Big(1-\frac{2M}{r}\Big)\frac{2M}{r^2}\mcM_1\Big\}\Big]
\ee
To summarize, we have fixed $\mcM_1$ \eqref{M1} in terms of the background solution, so that $K_{lm}$ behaves as $\mathcal{O}(1/r)$ asymptotically, and we have fixed other two gauge parameters, $\mcM_0, \mcM_2$, in terms of $\mcM_1$.
Finally we get for even parity perturbation in Zerilli gauge considering only  quadrapole perturbation($l=2$) and choosing $m=0$ as 
\be
h^{e}_{\mu\nu}=\begin{pmatrix}
H^{(N)}(t,r)(1-\frac{2M}{r})Y_2^0 & H^{(N)}_1(t,r)Y_2^0 & 0 & 0 \\
H^{(N)}_1(t,r)Y_2^0 & H^{(N)}(t,r)(1-\frac{2M}{r})^{-1}Y_2^0 & h^{(e)(N)}_{1}\pr_\theta{Y_2^0} & 0\\
0 & h^{(e)(N)}_{1}\pr_\theta{Y_2^0} & r^2\mathcal{T}_2^0 & 0\\
0 & 0 & 0 & r^2\sin^2\theta\mathcal{T}_2^0
\end{pmatrix}
\ee
where, 
$\mathcal{T}_2^0=K^{(N)}(t,r)Y_2^0+G^{(N)}(t,r)\pr^2_\theta{Y_2^0}$\\
and
$\tilde{\mathcal{T}}_2^0=K^{(N)}(t,r)Y_2^0+\cot\theta{G}^{(N)}(t,r)\pr_\theta{Y_2^0}$\\
Keep in mind that the terms containing $\pr_\phi{Y_2^0}(=0)$ has been omitted.
We list here the non zero components 
\be
H^{(N)}(t,r)={H_{lm}}(t,r)-2\Big(1-\frac{2M}{r}\Big)^{-1}\Big(\frac{\pr \mcM_0}{\pr t}-\frac{M}{r^3}(r-2M)\mcM_1\Big)
\ee
\be
H^{(N)}_1(t,r)={H_{1lm}}(t,r)-\Big(\frac{\pr \mcM_1}{\pr t}+\frac{\pr \mcM_0}{\pr r}-\frac{2M}{r(r-2M)}\mcM_0\Big)
\ee
\be
h^{(e)(N)}_1=-\Big(\frac{\pr \mcM_2}{\pr r}-\frac{2}{r}\mcM_2+\mcM_1\Big)
\ee
\be
K^{(N)}=K_{lm}(t,r)-\frac{2}{r^2}(r-2M)\mcM_1
\ee
\be
G^{(N)}_{lm}(t,r)=-\frac{2}{r^2}\mcM_2
\ee
One can check the trace of the perturbation behaves as
\be
h^\mu_\mu{\sim}[K^{(N)}-(\lambda+1)G^{(N)}]\sim\mathcal{O}\Big(\frac{1}{r^3}\Big)
\ee
near $r\to\infty$. Finally putting together the metric corresponding to even and odd parity perturbation and adding it with the ordinary Schwarzschild metric we have obtained the metric describing the ringing Schwarzschild black hole.  We have dropped the $``N"$ indices of the metric components when using in Sec.\ref{metric}. 
\section{Expression of the source terms ${\cal P}_{lmc\gamma}$ and $\bar{\cal P}_{lmc\gamma}$ of the main text}\label{Source_terms}
Thanks to our adopted perturbative approach which helps us 
use the separation of variable as $\xi^{lm}_{1}(t,r){\to}e^{-\mi(\omega+\km){t}}\tilde{\xi}^{{\km}lm}_{1}(r)$. Using this, the first equation of Eq.\eqref{part.sceq} can be transformed into 
\be
\bea
&f(r)\frac{1}{r^2}\pr_r\{r^2f(r)\pr_r\tilde{\xi}^{{\km}lm}_{1}(r)\}+\Big\{(\km+\omega)^2-f(r)\frac{l(l+1)}{r^2}\Big\}\tilde{\xi}^{{\km}lm}_{1}(r)+\frac{1}{2}P_{{\km}lm}(r)=0
\eea
\ee
where we see that the term, $\sum_{c\gamma} {\cal P}_{lmc\gamma} \xi^{c\gamma}_0=\frac{1}2{}e^{-\mi(\km+\omega){t}}P_{{\km}lm}(r)$, with
\be
\bea
&P_{{\km}lm}(r)\\
&=f(r)\sum_{c\gamma}\Lambda^{(2,0)}_{c\gamma{lm}}\Big[f(r)^{-1}(\omega\km+\km^2)(H^{(N)}+K^{(N)})\tilde{\xi}^{c\gamma}_0(r)+\frac{1}{r^2}\pr_r\{r^2f(r)(-H^{(N)}+K^{(N)})\pr_r\tilde{\xi}^{c\gamma}_0(r)\}\\
&-3f(r)^{-1}(\omega\km+\km^2)G^{(N)}\tilde{\xi}^{c\gamma}_0(r)-3\frac{1}{r^2}\pr_r\{r^2f(r)G^{(N)}\pr_r\tilde{\xi}^{c\gamma}_0(r)\}-\mi(\omega+\km)H^{(N)}_1\pr_r\tilde{\xi}^{c\gamma}_0(r)\\
&-\frac{1}{r^2}\mi\km\pr_r(r^2H^{(N)}_1\tilde{\xi}^{c\gamma}_0(r))+6\frac{1}{r^2}f(r)h^{(N)(e)}_1\pr_r\tilde{\xi}^{c\gamma}_0(r)-\frac{c(c+1)}{r^2}G^{(N)}\tilde{\xi}^{c\gamma}_0(r)\Big]\\
&+f(r)\frac{1}{r^2}\Big[\sum_{c\gamma}2\gamma\Lambda^{(2,0)}_{c\gamma{lm}}+3\sqrt{\frac{2}{3}}\sum_{c\gamma}\sqrt{(c-\gamma)(c+\gamma+1)}\Lambda^{(2,-1)}_{c(\gamma+1)lm}\Big]\times\\
&\times\Big[\pr_r\{f(r)h^{(N)(e)}_1\tilde{\xi}^{c\gamma}_0(r)\}+f(r)h^{(N)(e)}_1\pr_r\tilde{\xi}^{c\gamma}_0(r)-G^{(N)}\tilde{\xi}^{c\gamma}_0(r)\Big]\\
&+f(r)\frac{3}{2}\sqrt{\frac{5}{\pi}}\frac{1}{r^4}h^{(N)}_2\Big[\sqrt{\frac{\pi}{3}}\sum_{c\gamma}2\gamma\Lambda^{(1,0)}_{c\gamma{lm}}+2\sqrt{\frac{2\pi}{3}}\sum_{c\gamma}\Lambda^{(1,-1)}_{c(\gamma+1)lm}\sqrt{(c-\gamma)(c+\gamma+1)}\Big]\mi{\gamma}\tilde{\xi}^{c\gamma}_0(r)\\
&+f(r)\frac{1}{r^2}\sum_{c\gamma}\mi{\gamma}\sqrt{15}\Lambda^{(1,0)}_{c\gamma{lm}}\Big[\pr_r\{f(r)h^{(N)(o)}_1\tilde{\xi}^{c\gamma}_0(r)\}+f(r)h^{(N)(o)}_1\pr_r\tilde{\xi}^{c\gamma}_0(r)+\frac{1}{r^2}h^{(N)}_2\tilde{\xi}^{c\gamma}_0(r)\Big]\\
&+f(r)\frac{1}{r^2}\sqrt{\frac{5}{4\pi}}G^{(N)}l(l+1)\tilde{\xi}^{lm}_0(r)-f(r)\frac{1}{r^2}m^2G^{(N)}\frac{3}{2}\sqrt{\frac{5}{\pi}}\tilde{\xi}^{lm}_0(r)\\
&+f(r)\frac{1}{r^2}\frac{m}{2}\sqrt{\frac{5}{\pi}}\Big[\pr_r\Big\{f(r)h^{(N)(e)}_1\tilde{\xi}^{lm}_0(r)\Big\}+f(r)h^{(N)(e)}_1\pr_r\tilde{\xi}^{lm}_0(r)-G^{(N)}\tilde{\xi}^{lm}_0(r)\Big].
\eea
\ee
The $\Lambda^{(l',m')}_{lmc\gamma}$ is related to Wigner 3-jm symbol, orginating from the following spherical harmonics identity, 
\be
\bea
Y_l^m(\theta,\phi)Y_{l'}^{m'}(\theta,\phi)=\sum_{c\gamma}\Lambda^{(l',m')}_{lmc\gamma}Y_c^\gamma(\theta,\phi) .
\eea
\ee
Similarly for the other part of the perturbative solution, $\bar{\xi}^{lm}$, we will have associated operator $\bar{\cal P}_{lmc\gamma}$ which is the function of the complex conjugate of metric fluctuation, $h^*_{\mu\nu}$. For this case $\sum_{c\gamma} \bar{\cal P}_{lmc\gamma} \xi^{c\gamma}_0=\frac{1}2{}e^{-\mi(\km-\omega^*){t}}\bar{P}_{{\km}lm}(r)$, where $\bar{P}_{{\km}lm}(r)$ can be obtained by replacing $\omega \rightarrow -\omega^*$ in the expression of $P_{{\km}lm}(r)$ and simultaneoulsy taking the complex conjugate of the radial part of the fluctuation components.
\section{Calculation Of Absorption Cross Section} \label{cal_abs}
According to the construction described in Sec.\ref{def_abs}, we have considered the space-time outside the interaction surface to be static Schwarzschild. This particular fact enables us to define the conserved quantity associated with the energy momentum tensor at the asymptotic infinity. We have used the outgoing null coordinate to maintain the causality condition in our calculation, specifically during the identification of the ingoing part by matching the scalar field solution and its $r$-derivative at spatial infinity.
We shall now briefly discuss the procedure to obtain the energy flux in $(u,r)$ coordinate, with $u=t-r_*$, in which  Schwarzschild metric is given by
\be
ds^2=-f(r)du^2-2dudr+r^2d\Omega 
\ee
From the conservation law,
\be
\nabla_\mu{P^\mu}=0,
\ee
where $P^{\mu}={\mathcal{T}^\mu}_\nu\xi^\nu$, ${\mathcal{T}^\mu}_\nu$ is the energy-momentum tensor and $\xi^\nu$ is the Killing vector, we obtain the conserve quantity as
\be
\mathcal{E}=\int{d^3x}\sqrt{-g}P^0 .
\ee
Taking time derivative on both side we get
\be
\pr_u\mathcal{E}=\int{d^3x}\pr_u(\sqrt{-g}P^0)=-\int{d^3x}\pr_i(\sqrt{-g}P^i)
\ee
Now choosing a r-constant hyper surface will lead to 
\be
\bea
\pr_u\mathcal{E}&=-\int{r^2d\Omega}P^r\\
&=-\int{r^2d\Omega}{\mathcal{T}^r}_u\xi^u\\
&=-\int{r^2d\Omega}{\mathcal{T}^r}_u\\
&=-\int{r^2d\Omega}g^{r\alpha}{\mathcal{T}_{\alpha{u}}}\\
&=-\int{r^2d\Omega}[-{\mathcal{T}_{uu}}+f(r){\mathcal{T}_{ru}}].
\eea
\ee
The energy flux measured by a stationary observer at $r\to\infty$, per unit time is
\be
\pr_u\mathcal{E}=\int{d}\Omega{r^2}[{\mathcal{T}_{ru}}-{\mathcal{T}_{uu}}] .
\ee
For massless scalar field stress energy tensor $\mathcal{T}_{\mu\nu}$ is given by
\be
\mathcal{T}_{\mu\nu}=\frac{1}{2}(\pr_\mu\phi^*\pr_\nu\phi+\pr_\mu\phi\pr_\nu\phi^*)-\frac{1}{2}g_{\mu\nu}\pr^\alpha\phi^*\pr_\alpha\phi
\ee
where scalar field is expanded as $\phi=\sum \mathcal{N}_{\km{lm}}\xi^{lm}(u,r)Y_{lm}(\theta,\phi)$ with  $\mathcal{N}_{\km{lm}}$ being constant. We have solved the scalar field modes starting from the black hole horizon to the interaction surface $(r_{int})$ with an ingoing boundary condition. For this, the ringing Schwarzschild metric is playing as a source. Then using that solution as the time dependent initial condition at the interaction surface, we solve the same mode equation in the static Schwarzschild background. This procedure renders it difficult to identify the plane wave component at the asymptotic infinity. 
Therefore, in order to identify the incoming scalar wave propagating along `z-direction, we first use the well known Rayleigh expansion of the plane wave in terms of the partial wave as
\be
e^{-\mi\km(u+z_*)}e^{-\mi\km{z_*}}\sim\sum{e^{-\mi\km(u+r_*)}}\frac{e^{-\mi\km{r_*}}}{2\mi\km{r}}\sqrt{4\pi(2l+1)}Y_l^0(\Omega)+ \mbox{Outgoing}
\ee
where $z_*=r_*\cos\theta$ (recall that we have transformed, $t\to{u+z_*}$). Considering the asymptotic solution Eq.\eqref{asymp}, we first choose the time independent normalization condition as,
\be\label{plane.w.exp}
\mathcal{N}_{\km{lm}}=\frac{\sqrt{4\pi(2l+1)}}{2\mi\km{\mathcal{A}^{\km}_{{l}m}(u\to\infty,r_{int})}}\delta_m^0 .
\ee
Using this normalization factor, we approximately define  following incoming scalar wave of momentum $\km$ propagating along z-direction, 
\be
\phi^{\km}_{in}\sim\frac{\int\sum \mathcal{N}_{\km{lm}}\mathcal{A}_{{l}m}(u,\km, \omega)Y_{lm}(\theta,\phi)\sin\theta{d\theta}}{\int\sum\sqrt{4\pi(2l+1)}Y_l^0(\Omega)\sin\theta{d\theta}}2\mi{\km}e^{-\mi\km(u+z_*)}e^{-\mi\km{z_*}} .
\ee
The time dependent amplitude is defined in such a way that in the $u\to{\infty}$ limit, it reduces to unity. 
With this approximated  $\phi_{in}$, we obtain the total rate of ingoing flux for a given mode $\km$ as,
\be
\pr_u\mathcal{E}^{\km}_{in}=\int{d}\Omega{r^2}[{\mathcal{T}^{in}_{ru}}-{\mathcal{T}^{in}_{uu}}].
\ee
With all these ingredients we define the total absorption cross-section for every individual mode $(k)$ as
\be
\sigma^k_{ring}(u,r_{int}) = \frac {\pr_u\mathcal{E}^{\km}}{\pr_u\mathcal{E}^{\km}_{in}} = \sum_{l} \frac{\pr_u\mathcal{E}^{\km l}}{\pr_u\mathcal{E}^{\km}_{in}} =\sum_l \sigma^{\km l}_{ring}(u,r_{int})
\ee
We have discussed our numerical results of $\sigma^{\km l}_{ring}(u,r_{int})$, which is the partial absorption cross section.

%
\section{Time varying rotation of the plane of polarization of  photon}\label{photon_axion}
In this section we will study the  conversion of axion to photon (vise-versa), specifically focussing on the axion background. Although to start with we will consider a constant magnetic field along with the time dependent axion background. In the later part we will only consider the axion part only, having additional effects in the final results. Taking the  following action, where axion couples ($g_{\phi\gamma\gamma}$ is the coupling constant) with photon,  
\be
S=\int \sqrt{-g}d^4x[-\frac{1}{2}(\pr_\mu\phi\pr^\mu\phi+m^2_\phi\phi^2)-\frac{1}{4}F_{\mu\nu}F^{\mu\nu}-\frac{1}{4}g_{\phi\gamma\gamma}\phi F_{\mu\nu}\tilde{F}^{\mu\nu}]
\ee
we obtain the inhomogenous Maxwell equation by varying the action with respect to $A_\mu$ as,
\be\label{em.eq}
\bea
&\pr_\mu F^{\mu\nu}+g_{\phi\gamma\gamma}\tilde{F}^{\mu\nu}\pr_\mu\phi=0
\eea
\ee
and inhomogenous scalar equation by varying the action with respect to $\phi$ as,
\be\label{scalar.eq}
\pr_\mu\pr^\mu\phi-m^2_\phi\phi-\frac{1}{4}g_{\phi\gamma\gamma}F_{\mu\nu}\tilde{F}^{\mu\nu}=0
\ee
Considering the radiation gauge, $\bar{\nabla}\cdot\bar{A}=0,A_0=0$, both the scalar \eqref{scalar.eq} and Maxwell equation \eqref{em.eq} can be simplified as, 
\be
\bea
&\pr_t \bar{E}-\bar{\nabla}\times\bar{B}-g_{\phi\gamma\gamma}\bar{B}\pr_t\phi-g_{\phi\gamma\gamma}\bar{\nabla}\phi\times\bar{E}=0  \\
&-\pr^2_t\phi+\nabla^2\phi-m^2_\phi\phi+g_{\phi\gamma\gamma}\bar{B}\cdot\pr_t\bar{A}=0
\eea
\ee
Symmetry of the background-fields motivate us to consider the propagation direction of the axion and photon fluctuation along $z$, for  simplification. So, expressing the scalar and em field as background with fluctuation, 
\be
\bea
&\phi(t,z)=\phi_0(t)+\tilde{\phi}(t,z)\\
&\bar{B}=\bar{B}_0+\bar{\nabla}\times \bar{A}(t,z)
\eea
\ee
we can derive the linearized  equation governing the evolution of fluctuation as,
\be
-\pr^2_t\bar{A}+\nabla^2\bar{A}-g_{\phi\gamma\gamma}[\bar{B}_0\pr_t\tilde{\phi}(t,z)+\pr_t\phi_0(t)\bar{\nabla}\times\bar{A}]=0
\ee
and
\be
-\pr^2_t\tilde{\phi}+\nabla^2\tilde{\phi}-m^2_\phi\tilde{\phi}+g_{\phi\gamma\gamma}\bar{B}_0\cdot\pr_t\bar{A}=0
\ee
Assuming the time scale of variation of background field is much larger than the axion-photon wavelength. Also we consider $\omega\sim\km$, so that we can write the fluctuation as plane waves like
\be
\bea
&\tilde{\phi}(t,z)=\phi_\omega(t)e^{-\mi\omega t}e^{-\mi\omega z}\\
&A(t,z)=\mi\begin{pmatrix}
A^x_\omega(t)e^{-\mi\omega t}\\ A^y_\omega(t)e^{-\mi\omega t}
\end{pmatrix}e^{-\mi\omega z}
\eea
\ee
We obtain the fluctuation equation of the fourier modes of axion and photon, like Schrodinger equation as,
\be
\bea
&\mi\pr_t A^x_\omega(t)=-\frac{1}{2}g_{\phi\gamma\gamma}B^x_0\phi_\omega(t)+\frac{\mi}{2}g_{\phi\gamma\gamma}\dot\phi_0(t)A^y_\omega(t)\\
&\mi\pr_t A^y_\omega(t)=-\frac{1}{2}g_{\phi\gamma\gamma}B^y_0\phi_\omega(t)-\frac{\mi}{2}g_{\phi\gamma\gamma}\dot\phi_0(t)A^x_\omega(t)\\
&\mi\pr_t\phi_\omega(t)=\frac{m^2_\phi}{2\omega}\phi_\omega(t)-\frac{1}{2}g_{\phi\gamma\gamma}B^x_0A^x_\omega(t)-\frac{1}{2}g_{\phi\gamma\gamma}B^x_0A^y_\omega(t)
\eea
\ee
In matrix form these equations can be assembled as
\be\label{meq}
\mi\pr_t \begin{pmatrix}
\phi_\omega(t)\\
A^x_\omega(t)\\
A^y_\omega(t)
\end{pmatrix}=\begin{pmatrix}
\Delta_m & \Delta_x & \Delta_y \\
\Delta_x & 0 & \Delta_\phi\\
\Delta_y & -\Delta_\phi & 0   
\end{pmatrix}\begin{pmatrix}
\phi_\omega(t)\\
A^x_\omega(t)\\
A^y_\omega(t)
\end{pmatrix}
\ee 
where $\Delta_m=\frac{m^2_\phi}{2\omega}$, $\Delta_i=-\frac{1}{2}g_{\phi\gamma\gamma}B_i (i\to x,y)$, $\Delta_\phi=\frac{\mi}{2} g_{\phi\gamma\gamma}\dot{\phi}_0$. Considering $\Psi(t)=\{\phi_\omega(t), A^x_\omega(t), A^y_\omega(t)\}$, we rewrite the above equation 
\be\label{eom.a}
\mi\pr_t \Psi =[H_0+\tilde{H}(t)]\Psi
\ee
where ,
\be
\tilde{H}(t)=\begin{pmatrix}
0 & \Delta_x & \Delta_y \\
\Delta_x & 0 &\Delta_\phi\\
\Delta_y & -\Delta_\phi & 0   
\end{pmatrix}
\ee
In the following discussion we will work in interaction picture 
\be
\Psi_{int}(t)=\mathcal{U}^\dagger(t)\Psi(t),~~H_{int}=\mathcal{U}^\dagger(t)\tilde{H}(t)\mathcal{U}(t), ~~\mathcal{U}(t)=e^{-\mi\int^t H_0(t')dt'}
\ee
so that the equation \eqref{eom.a} becomes 
\be
\bea
&\mi\pr_t \Psi_{int}(t)=H_{int}\Psi_{int},\\
&\implies \Psi_{int}(t)=e^{-\mi\int^t H_{int}(t')dt'}\Psi_{int}(0)\implies\Psi^{n+1}_{int}(t)=-\mi\int^t H_{int}(t')dt'\Psi^n_{int}(t')
\eea
\ee
Taking upto second order (considering the coupling $g_{\phi\gamma\gamma}$ very small) we get 
\be
\Psi_{int}=\Big(1-\mi\int^t_{t_0} H_{int}(t')dt'-\int^t_{t_0} H_{int}(t')dt'\int^{t'}_{t_0} H_{int}(t'')dt''\Big)\Psi(0)
\ee
After substituting this expression we get 
\be
\bea
A_{x}(t)&=A_x(0)-\mi\Delta_{x}\tilde{\Delta}_{m}\phi^\omega(0)-\mi\int^t \Delta_\phi dt'\, A^\omega_y(0)-\int^t dt'\Delta_\phi \int^{t'} dt^{''} \tilde{\Delta}_m \Delta_x\phi^\omega(0)\\
&-\int^t dt' \Big(\Delta_x\tilde{\Delta}_{m}\int^{t'}dt^{''}\Delta_x\tilde{\Delta}_{m}-\Delta_\phi\int^{t'}dt^{''}\Delta_\phi\Big) A^\omega_x(0)-\int^t dt' \Delta_x\tilde{\Delta}_{m}\int^{t'}dt^{''}\Delta_y\tilde{\Delta}_{m} A^\omega_y(0)\\
A_{y}(t)&=A_y(0)-\mi\Delta_{y}\tilde{\Delta}_{m}\phi^\omega(0)+\mi\int^t \tilde{\Delta}_\phi dt' A^\omega_x(0)+\int^t dt'\Delta_\phi \int^{t'} dt^{''} \tilde{\Delta}_m \Delta_y\phi^\omega(0)\\
&-\int^t dt' \Big(\Delta_y\tilde{\Delta}_{m}\int^{t'}dt^{''}\Delta_y\tilde{\Delta}_{m}-\Delta_\phi\int^{t'}dt^{''}\Delta_\phi\Big) A^\omega_y(0)-\int^t dt' \Delta_y\tilde{\Delta}_{m}\int^{t'}dt^{''}\Delta_x\tilde{\Delta}_{m} A^\omega_x(0)
\eea
\ee
where, we have used $\Psi_{int}(0)=\Psi(0)$ and $\tilde{\Delta}_m=\int^t{\Delta}_m dt' $\eqref{meq}. Upto second order in the perturbative evaluation of stokes parameters we have found that the magnetic field background contributes separately with the axion background. We will consider the contribution coming solely from axion background.  So, without magnetic field background. We have found the expression of Stokes parameters (considering $\Delta_x=0=\Delta_y$) 
\be
\bea
&I(t)=I(0)\Big\{1+\tilde{\Delta}^*_\phi\tilde{\Delta}_\phi+\int^t dt' (\Delta_\phi\tilde{\Delta}_\phi+c.c)\Big\}+(\tilde{\Delta}_\phi+\tilde{\Delta}^*_\phi) V(0)\\
&Q(t)=Q(0)\Big\{1-\tilde{\Delta}^*_\phi\tilde{\Delta}_\phi+\int^t dt' (\Delta_\phi\tilde{\Delta}_\phi+c.c)\Big\}-\mi(\tilde{\Delta}_\phi -\tilde{\Delta}^*_\phi)U(0)\\
&U(t)=U(0)\Big\{1-\tilde{\Delta}^*_\phi\tilde{\Delta}_\phi+\int^t dt' (\Delta_\phi\tilde{\Delta}_\phi+c.c)\Big\}+\mi Q(0)(\tilde{\Delta}_\phi-\tilde{\Delta}^*_\phi)\\
&V(t)=V(0)\Big\{1-\tilde{\Delta}^*_\phi\tilde{\Delta}_\phi+\int^t dt' (\Delta_\phi\tilde{\Delta}_\phi+c.c)\Big\}+I(0)(\tilde{\Delta}_\phi+\tilde{\Delta}^*_\phi)\\
&Q(t)\pm\mi U(t)=\{1\mp (\tilde{\Delta}_\phi-\tilde{\Delta}^*_\phi)\}\langle Q(0)\pm\mi U(0)\rangle\sim e^{\mp\mi\theta}\langle Q(0)\pm\mi U(0)\rangle
\eea
\ee
It can be checked from the above expressions of stokes parameters that, $\phi$ being real, there will be no conversion between axion and photon. But different helicity states will be affected and time dependence of the axion field would lead to distinguishable effects. The rotation (up to first order in the coupling constant) of the plane of linear polarization, 
for very small $\theta$, can be identified as $\theta\sim (\tilde{\Delta}_\phi-\tilde{\Delta}^*_\phi)/\mi$. And the interesting point is that because of the ringing oscillation we will see time varying rotation of the linear polarization.

\end{document}